\documentclass[sigconf]{acmart}
\usepackage{algorithm}
\usepackage{algorithmic}
\usepackage{multirow}
\usepackage{colortbl}

\AtBeginDocument{%
  \providecommand\BibTeX{{%
    \normalfont B\kern-0.5em{\scshape i\kern-0.25em b}\kern-0.8em\TeX}}}

\copyrightyear{2021}
\acmYear{2021}
\setcopyright{acmcopyright}\acmConference[WI-IAT '21 Companion]{IEEE/WIC/ACM International Conference on Web Intelligence}{December 14--17, 2021}{ESSENDON, VIC, Australia}
\acmBooktitle{IEEE/WIC/ACM International Conference on Web Intelligence (WI-IAT '21 Companion), December 14--17, 2021, ESSENDON, VIC, Australia}
\acmPrice{15.00}
\acmDOI{10.1145/3498851.3498931}
\acmISBN{978-1-4503-9187-0/21/12}

\begin{document}

\title{Matching Social Issues to Technologies for Civic Tech by Association Rule Mining using Weighted Casual Confidence}

\author{Masato Kikuchi$^1$, Shun Shiramatsu$^1$, Ryota Kozakai$^2$, Tadachika Ozono$^1$}
\affiliation{%
 \department{Department of Computer Science, Graduate School of Engineering}
 \institution{Nagoya Institute of Technology}
 \city{Nagoya}
 \state{Aichi}
 \country{Japan}\\
  \institution{$^1$\{kikuchi, siramatu, ozono\}@nitech.ac.jp, $^2$ryouta039@gmail.com}
}

\renewcommand{\shortauthors}{Kikuchi et al.}
\renewcommand{\shorttitle}{Matching Social Issues to Technologies for Civic Tech using Weighted Casual Confidence}

\begin{abstract}
More than 80 civic tech communities in Japan are developing information technology (IT) systems to solve their regional issues.
Collaboration among such communities across different regions assists in solving their problems because some groups have limited IT knowledge and experience for this purpose.
Our objective is to realize a civic tech matchmaking system to assist such communities in finding better partners with IT experience in their issues.
In this study, as the first step toward collaboration, we acquire relevant social issues and information technologies by association rule mining.
To meet our challenge, we supply a questionnaire to members of civic tech communities and obtain answers on their faced issues and their available technologies.
Subsequently, we match the relevant issues and technologies from the answers. However, most of the issues and technologies in this questionnaire data are infrequent, and there is a significant bias in their occurrence.
Here, it is difficult to extract truly relevant issues--technologies combinations with existing interestingness measures.
Therefore, we introduce a new measure called weighted casual confidence, and show that our measure is effective for mining relevant issues--technologies pairs.
\end{abstract}




\maketitle

\section{Introduction}

Civic tech has been introduced in Europe~\cite{Lee:15}, wherein citizens in a region solve regional social issues using information technologies.
Since the Great East Japan Earthquake, many civic tech communities, which are civic groups that apply civic tech, have been established in various regions of Japan, and the number of communities is now over 80.
The issues that such communities deal with and their technical experience and knowledge are different for different communities.
However, some regions face similar or identical social issues.
For example, in the last two years, almost all regions have been facing problems related to COVID-19.
In such cases, the technical experience and knowledge of one community can be used for the issues that other communities are dealing with.
Moreover, cross-sectional collaboration across multiple regions is occasionally essential for solving problems.
Therefore, communities need to share their knowledge and assist each other in solving issues appropriately and rapidly.
To achieve collaboration among communities, the following information are necessary but insufficiently shared, which is a factor that inhibits collaboration.
\begin{itemize}
\item What technologies are necessary for solving a certain issue?
\item Which communities are the best at these technologies?
\end{itemize}
There are two reasons why information is difficult to share.
First, it is difficult to identify which technologies are useful for unresolved issues.
Hence, there are social technology officers (STOs), who are experts in matching issues and technologies.
However, hiring STOs is expensive for civic tech communities.
Second, it is difficult to determine the skill levels of all communities in the issues and technologies using human power because there are many communities in Japan.
Therefore, our final objective is to build a system that identifies the technologies required to solve issues and subsequently recommends the best communities for collaboration.

In this study, we match social issues with the information technologies required to solve them, as a first step toward the realization of the abovementioned recommendation system.
We supply a questionnaire to members of civic tech communities and obtain answers on their social issues and the information technologies they use.
However, the correspondence between the issues and the technologies is unclear from the answers.
Therefore, to match them appropriately, we use association rule mining to acquire such pairs based on their co-occurrence in the answers.

Most of the association rules generated from the answers are infrequent and need to be dealt with carefully.
There is also a significant bias in the occurrence of technologies, with many technologies being mentioned rarely, whereas a few technologies being specified in most answers.
To avoid these problems, we propose a novel measure called weighted casual confidence (WCC).
One approach to deal with the occurrence bias is using negative evidence.
Casual confidence~\cite{Kodratoff:01} was proposed as a measure that uses evidence.
However, because it treats positive and negative evidence equivalently, it mines many infrequent and unreliable rules.
In comparison, the proposed measure treats the two types of evidence with different weights.
Moreover, our measure uses conservative confidence~\cite{Kikuchi:16}, which underestimates the interestingness of a rule if it has a low frequency.
This allows dealing with low-frequency rules, which are ignored in general mining problems.
Consequently, our measure can mine issues--technologies pairs that frequently co-occur only with each other.
We conduct two experiments on mining results by supplying questionnaires to experts familiar with the technologies, and show the effectiveness of our measure in terms of the relevance of the issues and the technologies and their usefulness for knowledge discovery.

\section{Related Studies}

This section describes related studies on analyzing questionnaire data based on association rule mining and supporting collaboration of people.

Association rule mining has been used in various data analyses and for analyzing questionnaire data.
For example, Chen et al.~\cite{Chen:09} and Maduako et al.~\cite{Maduako:21} proposed mining approaches for dealing with different data types (e.g., open-ended answers, selective answers) of a questionnaire.
In our mining problem, as described in Section~\ref{sec:Questionnaire_data_for_civic_tech_communities}, we only use selective answers.
Specifically, we deal with only one data type.
Our questionnaire data form a transaction database, in which each transaction is a set of selective answers of a respondent.
Therefore, our mining problem can be formulated in the general association rule mining framework.
However, if we use the other answers of different data types for mining, we believe that these studies will be helpful.

Several types of research have been conducted to support collaboration, including civic tech communities.
Tossavainen et al.~\cite{Tossavainen:16} developed a matching system for social issues and social goals to support sustainable collaboration among civic tech communities.
Furthermore, Shiramatsu et al. ~\cite{Shiramatsu:15} conducted a practical research using this system in civic tech community events.
Horita et al.~\cite{Horita:17} studied designing a workshop to promote collaboration between researchers and citizens.
In comparison, we deal with a problem of matching different perspectives of social issues and information technologies for the objective of supporting collaboration.
In addition to the pairs of issues and technologies, the matching results of social issues and social goals by Tossavainen et al. may be used to provide better collaborative support.
Moreover, we believe that the matching of issues and technologies will be useful for supporting collaboration between researchers and citizens.

\section{Questionnaire Data for Civic Tech Communities}\label{sec:Questionnaire_data_for_civic_tech_communities}

\begin{table*}[tb]
\centering
\caption{Social issues. \#R denotes number of respondents who chose each issue.}
\label{tab:Issue_choices}
    \begin{tabular}{p{15em} r | p{15em} r | p{15em} r} \hline
        \multicolumn{6}{c}{Q: Please choose social issues your community is dealing with (multiple choice allowed).} \\ \hline
        \multicolumn{1}{c}{Social Issue} & \multicolumn{1}{c}{\#R} & \multicolumn{1}{| c}{Social Issue} & \multicolumn{1}{c}{\#R} & \multicolumn{1}{| c}{Social Issue} & \multicolumn{1}{c}{\#R}\\ \hline
        Agriculture & 5 & Forestry Industry & 2 & Raising Children & 20 \\
        Cooperation in Civic and Community & \multicolumn{1}{r |}{\multirow{2}{*}{28}} & \multicolumn{1}{l}{\multirow{2}{*}{Harbor}} & \multicolumn{1}{r |}{\multirow{2}{*}{0}} & \multicolumn{1}{l}{\multirow{2}{*}{Regional Archives}} & \multicolumn{1}{r}{\multirow{2}{*}{9}} \\
        Activities & & & & & \\
        COVID-19 & 26 & Health & 5 & River & 3 \\
        Crime Prevention & 2 & Housing & 0 & Road & 1 \\
        Cultural Exchange & 6 & Industry & 8 & SDGs & 17 \\
        Culture & 16 & Land & 0 & Sightseeing & 20 \\
        Disability Support & 6 & LGBTQ and Gender & 3 & Town Planning & 34 \\
        Disaster Prevention & 15 & Living & 11 & Traffic & 11 \\
        Education & 27 & Medical Care & 2 & Universe & 3 \\
        Elderly Support & 4 & Park & 5 & Upper and Lower Sewer & 0 \\
        Environment & 5 & Politics & 6 & Urban Planning & 6 \\
        Fishing Industry & 1 & Public and Private Partnership & 23 & Welfare & 5 \\
        Food & 13 & Public Facility & 5 & & \\
        \hline
    \end{tabular}
\end{table*}
\begin{table}[tb]
\centering
\caption{Information technologies. \#R denotes number of respondents who chose each technology.}
\label{tab:Tech_choices}
    \begin{tabular}{l r} \hline
        \multicolumn{2}{p{25em}}{Q: Please choose information technologies your community is using (multiple choice allowed).} \\ \hline
        \multicolumn{1}{c}{Information Technology} & \multicolumn{1}{r}{\#R} \\ \hline
        Chatbot & 5 \\ 
        Drone & 9 \\ 
        Game & 5 \\ 
        GIS and Geospatial Information & 26 \\ 
        Graphic Recording & 11 \\ 
        IoT & 19 \\ 
        Machine Learning & 11 \\ 
        No-code & 16 \\ 
        Open Data & 40 \\ 
        Programming & 27 \\ 
        Robot & 3 \\ 
        Smart Speaker & 3 \\ 
        SNS & 19 \\ 
        Telecommunications (5G, LoRa) & 2 \\ 
        Visualization & 18 \\ 
        Wikipedia and Wikidata & 16 \\ 
        \hline
    \end{tabular}
\end{table}

In our mining problem, we use a database based on the questionnaire data prepared in advance.
The questionnaire is designed to visualize the characteristics and specialties of the civic tech communities easily.
Although the questionnaire contains various questions, we use only the answers to the two questions mentioned in Tables~\ref{tab:Issue_choices} and \ref{tab:Tech_choices}.
We supply the questionnaire to 49 people from 47 civic tech communities in Japan and obtain answers about their social issues and the information technologies they use.
An example answer is expressed as
\begin{align*}
T_k =&\ \langle I_{issue}^k, I_{tech}^k \rangle \\
=&\ \langle \{\textrm{Sightseeing, Transportation, Living}\}, \\
&\{\textrm{Open Data, GIS, and Geospatial Information}\} \rangle,
\end{align*}
where $T_k$ represents an answer, and $k$ is an index that distinguishes the answers.
$I_{issue}^k$ and $I_{tech}^k$ are the sets of issues and technologies in $T_k$, respectively.
The range of $k$ is $1 \le k \le n$, and $n=49$ because 49 people answered the questionnaire.
In the above example, a community deals with three issues: ``sightseeing,'' ``transportation,'' and ``living,'' and uses two technologies: ``open data'' and ``GIS and geospatial information.''
In this paper, we denote a set of such answers as a database $D=\{T_k\}_{k=1}^{n}$ and each $T_k$ as a transaction.
In the next section, we present the problem formulation of matching high-relevance issues and technologies from their possible combinations in $T_k$s as association rule mining.
Accordingly, we obtain high-relevance issues--technologies pairs with statistical evidence.

\section{Formulation of Our Matching Problem}

From the questionnaire data described in the previous section, we match social issues with the information technologies required to solve them.
We formulate this problem as an association rule mining problem.
Association rule mining is a well-known data analysis method to obtain frequent patterns and combinations of high-relevance patterns in a transaction database.
By association rule mining, we can obtain high-relevance issues--technologies pairs based on statistical evidence from numerous combinations.

We represent an association rule as $X \Rightarrow Y$, where $X$ is a set of items related to the social issues and $Y$ is a set of items related to the information technologies.
The items are the answer choices listed in Tables~\ref{tab:Issue_choices} and~\ref{tab:Tech_choices} (e.g., ``COVID-19,'' ``cultural exchange,'' ``GIS and geospatial information'').
Suppose that the sets of all items for the issues and technologies are $I_{issue}$ and $I_{tech}$, respectively,
\begin{align*}
X \subseteq I_{issue},\quad Y \subseteq I_{tech}
\end{align*}
are satisfied.
The rule, $X \Rightarrow Y$, suggests that if the condition, $X$, is satisfied, then $Y$ is also satisfied.
For example, the following rule indicates that the technology of ``GIS and geospatial information'' is used for the issues of ``COVID-19'' and ``Traffic'':
\begin{align*}
{\rm \left\{COVID\mathchar`-19, Traffic\right\}} \Rightarrow \left\{\textrm{GIS and Geospatial Information} \right\}.
\end{align*}
Association rule mining identifies rules in which the relationship between $X$ and $Y$ is frequently satisfied (specifically, rules with strong ``interestingness'').
Interestingness measures, which are measures of the interestingness of rules, are described in Section~\ref{sec:Interestingness_measures}.

To perform association rule mining, we first enumerate all possible pairs $\langle X, Y\rangle$s of $X$ and $Y$.
For example, if $|I_{issue}|=|I_{tech}|=2$, there are $(\sum_{r=1}^{2}{}_2 {\rm C}_r)^2=9$ combinations of possible pairs $\langle X, Y \rangle$s.
In this example, it is easy to enumerate all pairs because there are only two choices each for the issues and the technologies.
However, in our case, as can be seen from Tables~\ref{tab:Issue_choices} and~\ref{tab:Tech_choices}, $|I_{issue}|=38$ and $|I_{tech}|=16$, and enumerating all pairs leads to a combinatorial explosion.
Therefore, we enumerate all pairs $\langle X, Y \rangle$s with $X$ occurring in two or more answers.
This reduces the total number of pairs to be enumerated to 7,733,793.
Note that in general association rule mining, a threshold is set for the frequency of pair $\langle X, Y \rangle$.
This threshold is extremely stringent to apply in our case.
Our threshold is set for the frequency of $X$ to enumerate many pairs, because our database is small and most of the rules are infrequent.

In Algorithm~\ref{alg:alg1}, the dataset, $D$, described in Section~\ref{sec:Questionnaire_data_for_civic_tech_communities} is the input, and its outputs are the issues--techniques pairs, $\langle X, Y \rangle$s, and their frequencies $cf(X)$, $cf(Y)$, and $cf(X \cup Y)$, which are the numbers of answers in which $X$, $Y$, and $X \cup Y$ occur, respectively.
${\rm Combi}(\cdot)$ is a function that takes a set as an argument and returns the set of all combinations among the elements.
The arguments, $I_{issue}^k$ and $I_{tech}^k$, are the sets of social issues and information technologies in the $k$-th answer $T_k$, respectively.
Therefore, $C_{issue}^k$ and $C_{tech}^k$ are the combinations of the issues and technologies generated from $T_k$, respectively.
For example, if $\textrm{Combi}(I_{tech}^k)$ and $I_{tech}^k=\left\{\textrm{Open Data, SNS}\right\}$, the returned value, $C_{tech}^k$, is as follows:
\begin{align*}
C_{tech}^k =&\ \{ \left\{ \textrm{Open Data} \right\}, \left\{\textrm{SNS} \right\}, \left\{ \textrm{Open Data, SNS} \right\} \}.
\end{align*}
This algorithm first enumerates the possible combinations among the issues and among the technologies from each answer $T_k$ in $D$, respectively.
Subsequently, it counts their frequencies $cf(X)$s and $cf(Y)$s.
Finally, it enumerates pairs $\langle X, Y \rangle$s that satisfy $cf(X) \ge 2$, and counts their co-occurrence frequencies $cf(X \cup Y)$s.
We use these frequencies to measure the interestingness of the association rules.

\begin{algorithm}                      
\caption{Generate pairs $\langle X, Y \rangle$s and count frequencies}         
\label{alg:alg1}
\begin{algorithmic}
\REQUIRE {Database $D = \{T_k\}_{k=1}^n$}
\ENSURE {$cf(X)$, $cf(Y)$, $cf(X \cup Y)$ for all $\langle X, Y \rangle \in R$}
\FORALL {transactions $T_k \in D$}
\STATE $C_{issue}^k \gets {\rm Combi}(I_{issue}^k),\ C_{tech}^k \gets {\rm Combi}(I_{tech}^k)$
\STATE Count frequency $cf(X)$ for all $X \in C_{issue}^k$
\STATE Count frequency $cf(Y)$ for all $Y \in C_{tech}^k$
\ENDFOR
\FOR {$k=1, 2, \ldots, n$}
\STATE Count frequency $cf(X \cup Y)$ for all $\langle X, Y \rangle \in R$, where
\STATE $R=\{\langle X, Y \rangle\ \mid cf(X) \ge 2,\ X \in C_{issue}^k,\ Y \in C_{tech}^k\}$
\ENDFOR
\end{algorithmic}
\end{algorithm}

\section{Interestingness Measures for Association Rules}\label{sec:Interestingness_measures}

Association rule mining measures the interestingness of each rule and mines the rules with strong interestingness.
Therefore, it is important to measure interestingness appropriately for successful mining.
In this section, we first introduce existing measures~\cite{Agrawal:94,Kodratoff:01,Kikuchi:16} and describe the problems of using them in our mining.
Subsequently, we propose the measure WCC to mitigate the problems.

\subsection{Confidence}

\begin{table}[tb]
\centering
\caption{Database example. $i_A, i_B, \cdots, i_G$ represent items.}
\label{tab:Database_example}
\begin{tabular}{l|ccccccc} \hline
Transaction 1 & $i_A$ & $i_B$ & & & $i_E$ & $i_F$ \\ \hline
Transaction 2 & $i_A$ & $i_B$ & & & $i_E$ & & $i_G$ \\ \hline
Transaction 3 & $i_A$ & $i_B$ & & & $i_E$ & & $i_G$ \\ \hline
Transaction 4 & $i_A$ & & & $i_D$ & & & $i_G$ \\ \hline
Transaction 5 & & $i_B$ & $i_C$ & & & $i_F$ & $i_G$ \\ \hline
Transaction 6 & & $i_B$ & $i_C$ & & & & $i_G$ \\ \hline
Transaction 7 & & $i_B$ & $i_C$ & & & & $i_G$ \\ \hline
Transaction 8 & & & $i_C$ & $i_D$ & & & $i_G$ \\ \hline
\end{tabular}
\end{table}

\begin{table*}[tb]
\centering
\caption{Estimation examples of the interestingness of association rules}
\label{tab:Estimation_example}
\begin{tabular}{ccccccccc} \hline
\multicolumn{1}{c}{Rule} & \multicolumn{4}{c}{Observed Frequency} & \multicolumn{4}{c}{Interestingness Measure} \\
$X \Rightarrow Y$ & $cf(X)$ & $cf(X \cup Y)$ & $cf(\overline{X})$ & $cf(\overline{X} \cup \overline{Y})$ &\multicolumn{1}{c}{Conf} & \multicolumn{1}{c}{${\rm Conf}_\ell$} & \multicolumn{1}{c}{Casual-Conf} & \multicolumn{1}{c}{WCC (Ours)} \\ \hline
$\{i_B\} \Rightarrow \{i_A\}$ & 6 & 3 & 2 & 1 & 0.500 & 0.142 & 0.100 & 0.125 \\
$\{i_F\} \Rightarrow \{i_A\}$ & 2 & 1 & 6 & 3 & 0.500 & 0.059 & 0.100 & 0.076 \\
$\{i_A\} \Rightarrow \{i_E\}$ & 4 & 3 & 4 & 4 & 0.750 & 0.222 & 0.310 & 0.275 \\
$\{i_A\} \Rightarrow \{i_G\}$ & 4 & 3 & 4 & 0 & 0.750 & 0.222 & 0.112 & 0.178 \\ \hline
\end{tabular}
\end{table*}

The classical association rule mining method, the Apriori algorithm ~\cite{Agrawal:94}, estimates the interestingness of a rule $X \Rightarrow Y$ as
\begin{align*}
{\rm Conf}(X \Rightarrow Y) = \frac{cf(X \cup Y)}{cf(X)},
\end{align*}
where $cf(X)$ is the number of transactions containing $X$ and $cf(X \cup Y)$ is the number of transactions containing both $X$ and $Y$.
Note that $X \cap Y=\emptyset$.
The above equation yields the maximum likelihood estimate of the conditional probability, $P(X \mid Y)$, which is called confidence.
Although confidence is easy to calculate, it can be unreasonably large for low-frequency rules.
For example, suppose we obtain the database summarized in Table~\ref{tab:Database_example}.
If we measure the interestingness of rules $\{i_B\} \Rightarrow \{i_A\}$ and $\{i_F\} \Rightarrow \{i_A\}$ with confidence, the estimates of both are 0.500, as listed in Table \ref{tab:Estimation_example}.
However, $\{i_A\}$ and $\{i_B\}$ co-occur thrice, whereas $\{i_A\}$ and $\{i_F\}$ co-occur only once, suggesting that the co-occurrence of $\{i_A\}$ and $\{i_F\}$ may be coincidental. 
Therefore, the Apriori algorithm introduces a threshold called minimum support ($minsup$) for the co-occurrence frequency and calculates the confidence for only those rules that satisfy $cf(X \cup Y) \ge minsup$.
In general association rule mining, the number of rules is much larger than in our case.
Therefore, by introducing $minsup$, we can reduce the number of rules to measure the interestingness and mitigate the problem of calculating the confidence for low-frequency rules.

\subsection{Conservative Confidence}

Our database consists of 49 answers (transactions).
Specifically, the maximum frequency that can be used for estimating the interestingness is 49, and most of the generated rules are infrequent.
In this case, it is difficult to determine the optimal $minsup$, and a small variation in $minsup$ will reduce numerous rules.
Therefore, Kikuchi et al.~\cite{Kikuchi:16} proposed a measure that weakly (conservatively) estimates the interestingness of a rule depending on its low frequency, instead of ignoring the low-frequency rules.
We call this measure ``conservative confidence'' (denoted as ${\rm Conf}_\ell$).
${\rm Conf}_\ell$ first constructs a confidence interval for $P(X \mid Y)$ using the method of Kikuchi et al.~\cite{Kikuchi:15}, and subsequently uses its lower bound as the interestingness score.
This measure has a confidence coefficient $1-\alpha$, where $0<\alpha<1$ is a parameter.
We set the coefficient as $0.99$ ($\alpha=0.01$) and use the lower bound of the one-sided 99\% confidence interval.
As listed in Table~\ref{tab:Estimation_example}, the scores for rules $\{i_B\}\Rightarrow\{i_A\}$ and $\{i_F\}\Rightarrow\{i_A\}$ are $0.142$ and $0.059$, respectively.
Therefore, using ${\rm Conf}_\ell$, we can preferentially identify high-frequency and reliable relationships and subsequently find low-frequency but frequently co-occurring relationships.
However, it is not always possible to obtain truly relevant relationships even using this measure.
Both rules $\{i_A\}\Rightarrow\{i_E\}$ and $\{i_A\}\Rightarrow\{i_G\}$ have $cf(X)=4$ and $cf(X \cup Y)=3$; therefore, both are estimated as 0.222.
However, item $i_E$ occurs only in the transactions that contain item $i_A$, whereas $i_G$ frequently occurs in the transactions that do not contain $i_A$.
Here, $i_A$ and $i_G$ may not be relevant.
However, ${\rm Conf}_\ell$ yields large estimates for such rules.

Table~\ref{tab:Tech_choices} shows that ``open data'' occurs in as many as 40 of the 49 answers.
Therefore, the use of ${\rm Conf}_\ell$ may result in the mining of many irrelevant issues--technologies pairs, including ``open data.''

\subsection{Casual Confidence}

One approach for mitigating the problem of ${\rm Conf}_\ell$ is using negative evidence (i.e., the frequency of the transactions in which the items do not occur).
Casual confidence~\cite{Kodratoff:01} was proposed as a measure for using negative evidence and is defined as 
\begin{align}
\label{eq:Casual-Conf}
{\rm Casual\mathchar`-Conf}(X \Rightarrow Y)=\frac{1}{2}\left[P(Y \mid X) + P(\overline{Y} \mid \overline{X})\right],
\end{align}
where $P(\overline{Y} \mid \overline{X})$ represents the probability that itemset $Y$ does not occur in a transaction under the condition that itemset $X$ does not occur in the transaction.
We estimate the two probabilities in the above equation using conservative confidence.
We set the confidence coefficients as $0.99$.
$P(\overline{Y} \mid \overline{X})$ is estimated from observed frequencies $cf(\overline{X})$ and $cf(\overline{X} \cup \overline{Y})$.
These frequencies can be obtained as
\begin{align*}
cf(\overline{X}) =&\ n-cf(X), \\
cf(\overline{X} \cup \overline{Y}) =&\ n-cf(X)-cf(Y)+cf(X \cup Y),
\end{align*}
respectively.
$n$ is the total number of transactions in the database.
As summarized in Table~\ref{tab:Estimation_example}, Casual-Conf can distinguish between rules $\{i_A\} \Rightarrow \{i_E\}$ and $\{i_A\} \Rightarrow \{i_G\}$, and the estimate of $\{i_A\} \Rightarrow \{i_E\}$ is larger than that of $\{i_A\} \Rightarrow \{i_G\}$.
Therefore, this measure can mitigate the problem encountered by ${\rm Conf}_\ell$.
However, similar to Conf, Casual-Conf yields the same estimates for rules $\{i_B\}\Rightarrow\{i_A\}$ and $\{i_F\}\Rightarrow\{i_A\}$.
Item $\{i_F\}$ occurs in only two transactions, and the estimate of $P(\overline{Y}\mid\overline{X})$ is as large as 0.142.
Casual-Conf yields a large estimate for infrequent rules because it is the arithmetic mean of $P(Y \mid X)$ and $P(\overline{Y} \mid \overline{X})$.
We treat many infrequent rules; therefore, using Casual-Conf results in mining almost only these rules.

\subsection{Weighted Casual Confidence}

We aim to acquire rule $X \Rightarrow Y$, where issues $X$ and technologies $Y$ frequently co-occur only with each other.
Therefore, instead of treating $P(Y \mid X)$ and $P(\overline{Y} \mid \overline{X})$ equally, we should pay attention to $P(Y \mid X)$.
To achieve this, we propose a measure that takes a weighted average of $P(Y \mid X)$ and $P(\overline{Y}\mid \overline{X})$, which is called WCC.
WCC is defined as
\begin{align}
\label{eq:WCC}
\textrm{WCC}(X \Rightarrow Y) = \frac{1}{2}\left[w \cdot P(Y \mid X) + (2-w) \cdot P(\overline{Y} \mid \overline{X})\right],
\end{align}
where $0<w<2$ is a weight parameter that adjusts the balance between $P(Y \mid X)$ and $P(\overline{Y} \mid \overline{X})$. 
When $w=1$, this measure is equivalent to Casual-Conf, as expressed in Eq.(\ref{eq:Casual-Conf}). 
In this study, we set $w=1.6$ to emphasize $P(Y \mid X)$. 
We estimate the two probabilities in the above equation using conservative confidence.
We set the confidence coefficients as $0.99$.
As summarized in Table~\ref{tab:Estimation_example}, the estimate of $\{i_B\} \Rightarrow \{i_A\}$ is larger than that of $\{i_F\} \Rightarrow \{i_A\}$.
Moreover, the estimate of $\{i_A\} \Rightarrow \{i_E\}$ is larger than that of $\{i_A\} \Rightarrow \{i_G\}$.
From the above, WCC can mitigate the problems encountered by both ${\rm Conf}_\ell$ and Casual-Conf.

\section{Experiments}

This section describes our experiments for identifying effective interestingness measures for matching the social issues and the information technologies.
Because there are no ground truth issues--technologies pairs, it is difficult to quantitatively evaluate the measures.
Hence, we conducted two experiments using questionnaires.
In the first experiment, we first performed association rule mining using three measures, except Conf, as described in Section~\ref{sec:Interestingness_measures}. Accordingly, we prepared three lists of high-scoring rules for each measure\footnote{Because it was noticeable from our experiments that Conf was ineffective, in cases where most of the rules were infrequent, we excluded this measure from the comparisons.}.
Subsequently, we supplied a questionnaire on these lists to approximately rank the measures.
Note that because ${\rm Conf}_\ell$ and WCC tend to assign similar scores to rules, the rule lists for these measures were expected to be similar.
Therefore, in the second experiment, we conducted a detailed comparison of the rules scored by ${\rm Conf}_\ell$ and WCC.
Specifically, we randomly showed a two-rule set to the respondents from the rule lists and subsequently supplied a questionnaire on the usefulness of the rules.
In this experiment, the number of rules we used and the number of respondents are larger than in the first experiment.

\subsection{Procedure for Experiment 1}

We showed the lists of rules generated for each measure to the respondents and supplied a questionnaire asking about the superiority or inferiority of the measures.
The objective of this experiment was to determine the approximate superiority or inferiority of the measures.
The experimental procedure was as follows.
First, we used Algorithm~\ref{alg:alg1} to generate the association rules and calculate their frequencies.
Next, we scored the rules using the three measures: ${\rm Conf}_\ell$, Casual-Conf, and WCC.
We extracted the top 30 rules with high scores for each measure, and listed these rules in descending order of scores.
Finally, we asked the respondents to compare these three lists and to answer the questions about the superiority and inferiority of the measures and their differences.
In this questionnaire, we hid the measure names and notated the measures as methods A, B, and C.
The 24 respondents consisted of research students, fourth-year undergraduate students, and graduate students majoring in computer science at a science and engineering university in Japan.

\subsection{Questionnaire for Experiment 1}

The questionnaire contained the following five questions:
\begin{itemize}
    \item[Q1-1:]Please rank methods A, B, and C.
You should choose a method like the following first: a method seems to generate the most plausible and useful pairs of social issues and information technologies
    \item[Q1-2:]Please indicate the difference between the first and second methods in terms of the relevance between the issues and technologies in each pair.
    \item[Q1-3:]Please indicate the difference between the first and second methods in terms of the usefulness for each pair of issues and technologies.
    \item[Q1-4:]Please indicate the difference between the second and third methods in terms of the relevance between the issues and technologies in each pair.
    \item[Q1-5:]Please indicate the difference between the second and third methods in terms of the usefulness for each pair of issues and technologies.
\end{itemize}
In this questionnaire, we asked the respondents to rank the measures and subsequently to indicate the differences in the superiority or inferiority of the measures.
They had the following four choices as the difference: (1) almost same, (2) slightly different, (3) different, or (4) largely different.

\subsection{Results of Experiment 1}

\begin{table}[tb]
\centering
\caption{Answers to the ranking of measures (Q1-1)}
\label{tab:Rank_of_methods}
\begin{tabular}{cccr} \hline
\multicolumn{3}{c}{Rank of Measures} & \multicolumn{1}{c}{\multirow{2}{*}{\#Respondents}} \\ \cline{1-3}
\multicolumn{1}{c}{1st (Best)} & \multicolumn{1}{c}{2nd} & \multicolumn{1}{c}{3rd (Worst)} &  \\ \hline
 ${\rm Conf}_\ell$ & Casual-Conf & WCC & 2 \\
 \multicolumn{1}{>{\columncolor[gray]{0.87}}c}{${\rm Conf}_\ell$} & \multicolumn{1}{>{\columncolor[gray]{0.87}}c}{WCC} & \multicolumn{1}{>{\columncolor[gray]{0.87}}c}{Casual-Conf} & \multicolumn{1}{>{\columncolor[gray]{0.87}}r}{8} \\
 Casual-Conf & ${\rm Conf}_\ell$ & WCC & 1 \\
 Casual-Conf & WCC & ${\rm Conf}_\ell$ & 1 \\
 \multicolumn{1}{>{\columncolor[gray]{0.87}}c}{WCC} & \multicolumn{1}{>{\columncolor[gray]{0.87}}c}{${\rm Conf}_\ell$} & \multicolumn{1}{>{\columncolor[gray]{0.87}}c}{Casual-Conf} & \multicolumn{1}{>{\columncolor[gray]{0.87}}r}{10} \\
 WCC & Casual-Conf & ${\rm Conf}_\ell$ & 2 \\ \hline
\end{tabular}
\end{table}

\begin{table*}[tb]
\centering
\caption{Answers to Q1-2 to Q1-5 (1st: ${\rm Conf}_\ell$, 2nd: $\rm WCC$, 3rd: $\rm Casual\mathchar`-Conf$)}
\label{tab:Q1-2_Q1-5_1}
\begin{tabular}{cclr|cclr} \hline
\multicolumn{8}{c}{Q1-2, Q1-4: Are the issues and technologies in each pair relevant?} \\ \hline
\multicolumn{2}{c}{Rank of Measures} & \multicolumn{1}{c}{\multirow{2}{*}{Choice (Q1-2)}} & \multicolumn{1}{c|}{\multirow{2}{*}{\#Respondents}} & \multicolumn{2}{c}{Rank of Measures} & \multicolumn{1}{c}{\multirow{2}{*}{Choice (Q1-4)}} & \multicolumn{1}{c}{\multirow{2}{*}{\#Respondents}} \\ \cline{1-2} \cline{5-6}
\multicolumn{1}{c}{1st} & \multicolumn{1}{c}{2nd} & & & \multicolumn{1}{c}{2nd} & \multicolumn{1}{c}{3rd} & &  \\ \hline
\multicolumn{1}{c}{\multirow{4}{*}{${\rm Conf}_\ell$}} & \multicolumn{1}{c}{\multirow{4}{*}{WCC}} & (1) almost same & 1 & \multicolumn{1}{c}{\multirow{4}{*}{WCC}} & \multicolumn{1}{c}{\multirow{4}{*}{Casual-Conf}} & (1) almost same & 1 \\
 & & (2) slightly different & 5 & & & (2) slightly different & 2 \\
 & & (3) different & 2 & & & (3) different & 5 \\
 & & (4) largely different & 0 & & & (4) largely different & 0 \\ \hline \hline
\multicolumn{8}{c}{Q1-3, Q1-5: Are the issues and technologies in each pair useful?} \\ \hline
\multicolumn{2}{c}{Rank of Measures} & \multicolumn{1}{c}{\multirow{2}{*}{Choice (Q1-3)}} & \multicolumn{1}{c|}{\multirow{2}{*}{\#Respondents}} & \multicolumn{2}{c}{Rank of Measures} & \multicolumn{1}{c}{\multirow{2}{*}{Choice (Q1-5)}} & \multicolumn{1}{c}{\multirow{2}{*}{\#Respondents}} \\ \cline{1-2} \cline{5-6}
\multicolumn{1}{c}{1st} & \multicolumn{1}{c}{2nd} & & & \multicolumn{1}{c}{2nd} & \multicolumn{1}{c}{3rd} & &  \\ \hline
\multicolumn{1}{c}{\multirow{4}{*}{${\rm Conf}_\ell$}} & \multicolumn{1}{c}{\multirow{4}{*}{WCC}} & (1) almost same & 6 & \multicolumn{1}{c}{\multirow{4}{*}{WCC}} & \multicolumn{1}{c}{\multirow{4}{*}{Casual-Conf}} & (1) almost same & 1 \\
 & & (2) slightly different & 1 & & & (2) slightly different & 1 \\
 & & (3) different & 1 & & & (3) different & 6 \\
 & & (4) largely different & 0 & & & (4) largely different & 0 \\ \hline
\end{tabular}
\end{table*}

\begin{table*}[tb]
\centering
\caption{Answers to Q1-2 to Q1-5 (1st: $\rm WCC$, 2nd: ${\rm Conf}_\ell$, 3rd: $\rm Casual\mathchar`-Conf$)}
\label{tab:Q1-2_Q1-5_2}
\begin{tabular}{cclr|cclr} \hline
\multicolumn{8}{c}{Q1-2, Q1-4: Are the issues and technologies in each pair relevant?} \\ \hline
\multicolumn{2}{c}{Rank of Measures} & \multicolumn{1}{c}{\multirow{2}{*}{Choice (Q1-2)}} & \multicolumn{1}{c|}{\multirow{2}{*}{\#Respondents}} & \multicolumn{2}{c}{Rank of Measures} & \multicolumn{1}{c}{\multirow{2}{*}{Choice (Q1-4)}} & \multicolumn{1}{c}{\multirow{2}{*}{\#Respondents}} \\ \cline{1-2} \cline{5-6}
\multicolumn{1}{c}{1st} & \multicolumn{1}{c}{2nd} & & & \multicolumn{1}{c}{2nd} & \multicolumn{1}{c}{3rd} & &  \\ \hline
\multicolumn{1}{c}{\multirow{4}{*}{WCC}} & \multicolumn{1}{c}{\multirow{4}{*}{${\rm Conf}_\ell$}} & (1) almost same & 4 & \multicolumn{1}{c}{\multirow{4}{*}{${\rm Conf}_\ell$}} & \multicolumn{1}{c}{\multirow{4}{*}{Casual-Conf}} & (1) almost same & 1 \\
 & & (2) slightly different & 6 & & & (2) slightly different & 4 \\
 & & (3) different & 0 & & & (3) different & 3 \\
 & & (4) largely different & 0 & & & (4) largely different & 2 \\ \hline \hline
\multicolumn{8}{c}{Q1-3, Q1-5: Are the issues and technologies in each pair useful?} \\ \hline
\multicolumn{2}{c}{Rank of Measures} & \multicolumn{1}{c}{\multirow{2}{*}{Choice (Q1-3)}} & \multicolumn{1}{c|}{\multirow{2}{*}{\#Respondents}} & \multicolumn{2}{c}{Rank of Measures} & \multicolumn{1}{c}{\multirow{2}{*}{Choice (Q1-5)}} & \multicolumn{1}{c}{\multirow{2}{*}{\#Respondents}} \\ \cline{1-2} \cline{5-6}
\multicolumn{1}{c}{1st} & \multicolumn{1}{c}{2nd} & & & \multicolumn{1}{c}{2nd} & \multicolumn{1}{c}{3rd} & &  \\ \hline
\multicolumn{1}{c}{\multirow{4}{*}{WCC}} & \multicolumn{1}{c}{\multirow{4}{*}{${\rm Conf}_\ell$}} & (1) almost same & 5 & \multicolumn{1}{c}{\multirow{4}{*}{${\rm Conf}_\ell$}} & \multicolumn{1}{c}{\multirow{4}{*}{Casual-Conf}} & (1) almost same & 1 \\
 & & (2) slightly different & 3 & & & (2) slightly different & 3 \\
 & & (3) different & 2 & & & (3) different & 4 \\
 & & (4) largely different & 0 & & & (4) largely different & 2 \\ \hline
\end{tabular}
\end{table*}

First, we focus on Table 5, which provides the answers to the ranking of the measures (Q1-1).
As listed in the column of the first rank, 12 and 10 respondents answered WCC and ${\rm Conf}_\ell$, respectively.
Thus, WCC is better than ${\rm Conf}_\ell$, with a difference of only two respondents.
In contrast, in the column of the third rank, 18 out of 24 respondents answered Casual-Conf, indicating that many determined it to be inferior to the other two measures.

Next, we focus on the differences among the measures. 
We received various answers; however, owing to space limitations, we cannot include all of them.
Therefore, we limit our discussion to the two most prominent choices in Q1-1 (highlighted in gray in Table~\ref{tab:Rank_of_methods}).
Tables~\ref{tab:Q1-2_Q1-5_1} and \ref{tab:Q1-2_Q1-5_2} list the answers of Q1-2 to Q1-5 for the two choices, respectively.
For each table, we first focus on the differences between the first and second measures.
These represent the differences between ${\rm Conf}_\ell$ and WCC, and the majority of respondents answered ``almost same'' and ``slightly different'' in terms of both relevance and usefulness.
Subsequently we focus on the differences between the second and third measures.
These represent the differences between ${\rm Conf}_\ell$ and Casual-Conf or the difference between WCC and Casual-Conf.
For the differences between Casual-Conf and ${\rm Conf}_\ell$ or WCC, the majority of respondents answered ``different'' in all cases, except Q1-4, based on Table~\ref{tab:Q1-2_Q1-5_2}.
From Table~\ref{tab:Q1-2_Q1-5_2}, we can see that two respondents answered ``largely different'' for the difference between ${\rm Conf}_\ell$ and Casual-Conf.

We summarize the results of the questionnaire.
In terms of the effectiveness of the measures in matching the social issues and the information technologies, many respondents answered that WCC or ${\rm Conf}_\ell$ as the best (12 and 10 respondents, respectively), whereas many answered Casual-Conf as the worst (18 respondents).
As for the differences among the measures, WCC and ${\rm Conf}_\ell$ were close, whereas Casual-Conf was considered inferior to these two measures by many respondents.

\subsection{Procedure for Experiment 2}

We randomly selected a two-rule set awarded high scores by ${\rm Conf}_\ell$ or WCC or both, showed it to the respondents, and asked them to compare the goodness or badness of the two rules.
We conducted experiments to clarify the difference in the superiority and inferiority of $ {\rm Conf}_\ell $ and WCC.
Because it was difficult for the respondents to provide an absolute rating for each rule, we asked them to assign a relative rating between the two rules.
The experimental procedure was as follows.
First, from the scored rules, we created two lists of 100 rules awarded high scores by ${\rm Conf}_\ell$ and WCC, respectively.
Next, we took the union of the rule lists.
Consequently, there were 63 rules contained in both lists, and after removal of duplication, the union size was 137.
Here, we assigned unique labels to the rules that were contained only in the list of ${\rm Conf}_\ell$ or WCC and both lists, respectively.
We used these labels only to analyze the answers and did not show them to the respondents.
Finally, we randomly selected a two-rule set with different labels from this union and showed to the respondents for comparison of their goodness or badness.
We conducted a crowdsourced questionnaire of people with experience in information technology and received 817 answers\footnote{We used Lancers (\url{https://www.lancers.jp/}) as the crowdsourcing service. The response period was six days.}.
Experience in information technology implied experience in learning information technology or engaging in work related to information technology.

\subsection{Questionnaire for Experiment 2}

Before the questionnaire, we explained in writing to the respondents that our final objective was to create a recommendation system for information technologies required for solving social issues.
Next, we presented the respondents with pair 1 $\langle X_1, Y_1 \rangle$ and pair 2 $\langle X_2, Y_2 \rangle$ of issues and technologies that constitute the two-rule set \{$X_1 \Rightarrow Y_1$, $X_2 \Rightarrow Y_2$\}, and asked them to compare the usefulness of these pairs in the recommendation system.

\begin{itemize}
    \item[Q2:]Please compare the usefulness of pairs 1 and 2 for the recommendation system and choose the appropriate one from the following choices:
\end{itemize}
\begin{itemize}
    \item[(1)]Both of them are useful. In addition, pair 1 is more useful.
    \item[(2)]Both of them are useful. In addition, pair 2 is more useful.
    \item[(3)]Pair 1 is as useful as pair 2.
    \item[(4)]Pair 1 is useful, but pair 2 is not.
    \item[(5)]Pair 2 is useful, but pair 1 is not.
    \item[(6)]Both of them are not useful.
\end{itemize}

\subsection{Results of Experiment 2}

\begin{table*}[tb]
\centering
\caption{Comparison results of two rules with different labels.
\{${\rm Conf}_\ell$, ${\rm WCC}$\} is label assigned to rule existing in both lists.}
\label{tab:Comparison_between_rules}
\begin{tabular}{cclrr} \hline
\multicolumn{2}{c}{Labels of Two Rules} & \multicolumn{1}{c}{Choice} & \multicolumn{1}{c}{\# Respondents} & \multicolumn{1}{c}{Ratio} \\ \hline
\multicolumn{1}{c}{\multirow{6}{*}{${\rm Conf}_\ell$}} & \multicolumn{1}{c}{\multirow{6}{*}{\{${\rm Conf}_\ell$, WCC\}}} & (1)\ \ Both of them are useful. In addition, ${\rm Conf}_\ell$ is more useful. & 53 & 0.171 \\
 &  & \multicolumn{1}{>{\columncolor[gray]{0.87}}l}{(2)\ \ Both of them are useful. In addition, \{${\rm Conf}_\ell$, WCC\} is more useful.} & \multicolumn{1}{>{\columncolor[gray]{0.87}}r}{104} & \multicolumn{1}{>{\columncolor[gray]{0.87}}r}{0.335} \\
 &  & (3)\ \ ${\rm Conf}_\ell$ is as useful as \{${\rm Conf}_\ell$, WCC\}. & 87 & 0.281 \\
 &  & (4)\ \ ${\rm Conf}_\ell$ is useful, but \{${\rm Conf}_\ell$, WCC\} is not. & 23 & 0.074 \\
 &  & (5)\ \ \{${\rm Conf}_\ell$, WCC\} is useful, but ${\rm Conf}_\ell$ is not. & 31 & 0.100 \\
 &  & (6)\ \ Both of them are not useful. & 12 & 0.039 \\ \hline
 \multicolumn{1}{c}{\multirow{6}{*}{${\rm Conf}_\ell$}} & \multicolumn{1}{c}{\multirow{6}{*}{WCC}} & (1)\ \ Both of them are useful. In addition, ${\rm Conf}_\ell$ is more useful. & 38 & 0.205 \\
 &  & \multicolumn{1}{>{\columncolor[gray]{0.87}}l}{(2)\ \ Both of them are useful. In addition, WCC is more useful.} & \multicolumn{1}{>{\columncolor[gray]{0.87}}r}{79} & \multicolumn{1}{>{\columncolor[gray]{0.87}}r}{0.427} \\
 &  & (3)\ \ ${\rm Conf}_\ell$ is as useful as WCC. & 43 & 0.232 \\
 &  & (4)\ \ ${\rm Conf}_\ell$ is useful, but WCC is not. & 11 & 0.059 \\
 &  & (5)\ \ WCC is useful, but ${\rm Conf}_\ell$ is not. & 11 & 0.059 \\
 &  & (6)\ \ Both of them are not useful. & 3 & 0.016 \\ \hline
 \multicolumn{1}{c}{\multirow{6}{*}{\{${\rm Conf}_\ell$, WCC\}}} & \multicolumn{1}{c}{\multirow{6}{*}{WCC}} & (1)\ \ Both of them are useful. In addition, \{${\rm Conf}_\ell$, WCC\} is more useful. & 79 & 0.245 \\
 &  & \multicolumn{1}{>{\columncolor[gray]{0.87}}l}{(2)\ \ Both of them are useful. In addition, WCC is more useful.} & \multicolumn{1}{>{\columncolor[gray]{0.87}}r}{109} & \multicolumn{1}{>{\columncolor[gray]{0.87}}r}{0.339} \\
 &  & (3)\ \ \{${\rm Conf}_\ell$, WCC\} is as useful as WCC. & 78 & 0.242 \\
 &  & (4)\ \ \{${\rm Conf}_\ell$, WCC\} is useful, but WCC is not. & 20 & 0.062 \\
 &  & (5)\ \ WCC is useful, but \{${\rm Conf}_\ell$, WCC\} is not. & 31 & 0.096 \\
 &  & (6)\ \ Both of them are not useful. & 5 & 0.016 \\ \hline
\end{tabular}
\end{table*}

Table~\ref{tab:Comparison_between_rules} summarizes the aggregate results of the answers based on the combinations of the labels for pairs 1 and 2.
The labels represent the measures for the lists that contained the pairs.
For ease of understanding the results, we replaced pairs 1 and 2 with the corresponding labels in the choices.
For each comparison between the labels, the most prominent choice, number of respondents who selected the choice, and percentage of the respondents for the six choices are highlighted in gray.

From the overall comparison results between the two labels, we can see that the percentages of both pairs being useful (i.e., the percentage of respondents who chose the top three choices among the six choices) is approximately 80\%.
In the comparison between ${\rm Conf}_\ell$ and \{${\rm Conf}_\ell$, WCC\}, the number of respondents who answered that ``\{${\rm Conf}_\ell$, WCC\} is more useful'' is approximately twice that who answered ``${\rm Conf}_\ell$ is more useful'' (104 and 53, respectively).
In the comparison between ${\rm Conf}_\ell$ and WCC, the number of respondents who answered that ``WCC is more useful'' is approximately twice that who answered that ``${\rm Conf}_\ell$ is more useful'' (79 and 38, respectively).
In the comparison between \{${\rm Conf}_\ell$, WCC\} and WCC, the number of respondents who answered that ``WCC is more useful'' is approximately 1.5 times that who answered that ``\{${\rm Conf}_\ell$, WCC\} is more useful'' (109 and 79, respectively).

We summarize the results of the questionnaire.
We asked whether the rules assigned high scores by ${\rm Conf}_\ell$ or WCC or both were useful for the recommendation system. 
Thus, approximately 80\% of the respondents answered that these rules were useful.
The rules awarded high scores only by WCC tended to be more useful than the other rules.
The rules assigned high scores by both ${\rm Conf}_\ell$ and WCC tended to be more useful than those only by ${\rm Conf}_\ell$.
Overall, we confirmed the effectiveness of WCC.

\section{Discussion}

In experiment 1, we could not observe noticeable differences in the superiority or inferiority of ${\rm Conf}_\ell$ and WCC.
We consider this is owing to the similarity of the rule lists of both measures.
We set the weight parameter, $w$, of WCC as 1.6.
In this setting, the scores of WCC are close to those of ${\rm Conf}_\ell$.
Consequently, of the 30 rules contained in the list of WCC, more than half (18 rules) are also contained in the list of ${\rm Conf}_\ell$.
Hence, we infer that this made it difficult for the respondents to differentiate between WCC and ${\rm Conf}_\ell$.
In contrast, the rules in the list of Casual-Conf do not overlap with the lists of the other rules lists.
However, the list contains only the rules that occur thrice, and the occurrence of these rules may be coincidental.
Therefore, Casual-Conf is determined to be inferior to the other two measures.

In experiment 2, the rules assigned high scores by only WCC tend to be more useful than the other rules.
To clarify the reason, we categorized the rules by their labels and compared them.
We found that more than 90\% of the rules labeled as ${\rm Conf}_\ell$ and \{${\rm Conf}_\ell$, WCC\} contained only ``GIS and geospatial information'' or ``open data'' or both in technologies $Y$.
These two technologies are chosen by numerous communities in the questionnaire, as shown in Section~\ref{sec:Questionnaire_data_for_civic_tech_communities}, and it is difficult to clarify their relationship with specific issues.
Therefore, we consider the rules awarded high scores by ${\rm Conf}_\ell$ are highly general and not useful.
However, the rules labeled as WCC contain technologies such as ``Wikipedia and Wikidata,'' ``IoT,'' ``visualization,'' ``SNS,'' ``programming,'' indicating that our proposed measure, WCC, can present more diverse rules than ${\rm Conf}_\ell$.
Based on the above, we consider that WCC is the superior measure.

As expressed in Eq.(\ref{eq:WCC}), WCC adjusts the impact of the two probabilities, $P(Y \mid X)$ and $P(\overline{Y} \mid \overline{X})$, on the estimates by the weight parameter, $w$.
In our experiments, we set $w$ as 1.6 to acquire positive association rules $X \Rightarrow Y$. 
However, if we set $w$ smaller than 1, WCC can acquire negative association rules $\overline{X}\Rightarrow\overline{Y}$.
In addition, WCC can deal with low-frequency rules without ignoring them by underestimating their interestingness.
However, WCC has two hyperparameters: confidence coefficient and weight $w$. 
Our future study will be on automatically setting these parameters depending on the mining objective and the properties of the rules.
Moreover, even WCC assigns high scores to the highly general relationships containing ``open data.''
Therefore, WCC can be further improved to eliminate these rules.
Furthermore, the effectiveness of WCC shown in our experiments is the result of the subjective evaluation achieved by questionnaires.
Although the rules we treated cannot be completely quantitatively evaluated for their correctness, it may be possible to evaluate quantitatively their correctness.
For example, by collecting source codes and related documents published by civic tech communities using crawlers and analyzing them, we may be able to link issues and technologies.
Quantitative evaluation of the tendency for the correctness of rules is necessary to support collaboration among communities, and is another future study.

\section{Conclusion}

In this paper, to support collaboration among civic tech communities, we presented a method to find high-relevance pairs of social issues and information technologies using questionnaire data on the issues faced by community members and their available technologies.
Based on the co-occurrence of issues and technologies in the questionnaire data obtained in advance, we formulated an association rule mining problem to separate the relevant issues and technologies.
However, most of the rules were infrequent, and there was a significant bias in the occurrence of some issues and technologies.
Here, ${\rm Conf}_\ell$~\cite{Kikuchi:16} cannot deal with the occurrence bias, and Casual-Conf~\cite{Kodratoff:01} mines infrequent and unreliable rules.
Therefore, we proposed WCC, which can mitigate both problems encountered by the above measures.
We conducted two experiments on mining using questionnaires.
In experiment 1, we showed the rule lists for each measure to 24 university students and asked them to rank the measures.
Many answered that WCC or ${\rm Conf}_\ell$ was the best (12 and 10 students, respectively), whereas 18 students answered that Casual-Conf was the worst.
In experiment 2, we showed a two-rule set assigned high scores by ${\rm Conf}_\ell$ or WCC or both to cloud workers and asked them to compare their usefulness for knowledge discovery.
Approximately 80\% of them answered that the two-rule set was useful.
In the comparisons of WCC with ${\rm Conf}_\ell$ and \{${\rm Conf}_\ell$, WCC\}, most of them answered that the rules of WCC were more useful than those of the others (79 and 109 workers, respectively).
In the comparison of \{${\rm Conf}_\ell$, WCC\} with ${\rm Conf}_\ell$, most of the respondents answered that the rules of \{${\rm Conf}_\ell$, WCC\} were more useful than those of ${\rm Conf}_\ell$ (104 workers).
Overall, we confirmed the effectiveness of WCC.

\begin{acks}
This work was supported in part by JSPS KAKENHI Grant Numbers JP19K12266, JP17K00461.
\end{acks}

\bibliographystyle{ACM-Reference-Format}
\bibliography{references_a}

\end{document}